\documentclass[prl,twocolumn]{revtex4}
\usepackage{graphicx}

\begin{document}
\title{Vortices in condensate mixtures}
\author{Christophe Josserand\\
\small  Laboratoire de Mod\'elisation en M\'ecanique\\
\small UPMC-CNRS UMR 7607, 4 place Jussieu, 75252 Paris C\'edex 05 France\\
Yves Pomeau\\
\small Laboratoire de Physique Statistique de l'Ecole
normale sup\'erieure, \\ 
\small 24 Rue Lhomond, 75231 Paris C\'edex 05, France
}

\begin{abstract}
    In a condensate made of two different atomic molecular 
    species, Onsager's quantization condition implies that around a vortex 
    the velocity field cannot be the same for the two species. We explore 
    some simple consequences of this observation. Thus if the two
    condensates are in slow relative translation one over the other, the 
    composite vortices are carried at a velocity that is a fraction 
    of the single species velocity.  This property is valid for attractive interaction and below
    a critical velocity which corresponds to a saddle-node bifurcation. 
\end{abstract}
\maketitle

One remarkable result in condensed matter physics is the 
discovery by Onsager\cite{onsag} that, in a superfluid, the vorticity can be 
present along narrow lines with a quantized circulation. Indeed the 
integral $\int\, \bf{u}.\bf{ds}$ of the fluid velocity ${\bf u}$ taken along a circuit enclosing a 
vortex line must be a positive or negative integer multiple of $\frac{h}{m}$, $h$ 
Planck's constant, $m$ mass of the particles making the superfluid. This 
 brings a striking analogy between the dynamics of the quantum 
 vortices of a superfluid and the Kelvin vortices of a classical inviscid fluid, the 
 quantization being present only to specify the value of the 
 circulation, an arbitrary quantity in a classical fluid. This 
 analogy between classical inviscid fluids and superfluids 
 is at the heart 
 of our understanding of superfluid mechanics, beginning with the 
 Landau two-fluid theory\cite{lan41} when there is no normal fluid. However 
 things are not so simple, just because the quantization 
 of the circulation involves explicitely the mass of the particles. 
 Therefore, if there is more than one species of particles with 
 different masses, it is not obvious at all that classical fluid mechanics 
 remains the right theory to describe the large scale motion of this 
 mixture with quantum vortices. This is because, in such a mixture, 
 one does not know {\it a priori} which mass enters into Onsager's 
circulation condition. 
 
 This question seems to be irrelevant for superfluid Helium 
 4, because it has no other stable bosonic isotope, and mixing it with 
any other atomic or molecular liquid is not possible at temperature low 
 enough to observe superfluidity. 
 Mixtures of Helium 4 and 3 (a Fermion) can remain liquid, although the 
 spin effects in Helium 3 make the whole picture quite different, but 
 certainly extremely interesting from the present point of view.  
 Below 
 we look at a situation that can be, presumably, realised in atomic 
 vapors, namely a mixture of two bosonic atoms (or eventually 
 molecules)\cite{Cornell,Hulet,Ketterle}. 
 
 We consider the following general problem: given that there are two 
 species in the same gas, both condensed, what are the dynamical properties of the 
 large scale motion of this mixture? This problem, as far we are 
 aware of, has not been looked at previously with vortices included ( without  vortex look at
 \cite{fetter,stringa}).

 By extrapolating what is 
 known in the case of single species condensate, one can think to 
 many relevant questions. The normal modes extending to mixtures the Bogoliubov spectrum
 has been studied by \cite{fetter}, while the density profiles in harmonic traps
 has been investigated by \cite{ho}. When looking  to the fluid motion itself, one of the most interesting issues there is the behaviour of vortices.
We assume that the mixture is at zero 
temperature and that each molecular/atomic species of molecular mass $m_j$ is described by a 
macroscopic wave function $\Psi_{j}({\bf{r}}, t)$, a complex valued 
function of the position ${\bf{r}}$ and of time $t$ with the discrete index  
$j$ being either $1$ or $2$, to denote the species under 
consideration (one could deal as well with more than two species). 
The equation of evolution of the coupled  $\Psi_{j}({\bf{r}}, t)$ 
$j = 1,2$ is {\it a priori} an extension\cite{fetter} of the familiar 
Gross-Pitaevskii equation (G-P later on): 
\begin{equation}
    i\hbar \frac{\partial \Psi_{j}}{\partial t} = 
    -\frac{\hbar^2}{2m_{j}}\nabla^2 \Psi_{j} + a_{j} |\Psi_{j}|^2  \Psi_{j} 
    + g  |\Psi_{(j+1)}|^2  \Psi_{j}
\label{eq1}
\end{equation}
The above writing is for two coupled equations, with $j = 1$ and $j = 
2$. In the interaction term, the index $(j+1)$ is computed {\it mod} 2: $1+1 = 2$, 
$2+1 = 1$.
Lastly the interaction real parameters $a_{j}$ and $g$ are such that the mixture is stable against 
collapse and against separation into 
two phases, one rich in $1$, the other in $2$. The stability 
depends on the minimum of the interaction part of the energy,
the volume integral of 
$$ \left(\frac{a_{1}}{2} |\Psi_{1}|^4 + \frac{a_{2}}{2} 
|\Psi_{2}|^4 + g |\Psi_{1}|^2 |\Psi_{2}|^2 \right) $$
The mixture is then stable against collapse if $a_1$ and $a_2$ are positive and
if $a_{1} a_{2} > g^2$. The linear stability against demixing 
is determined by the Bogoliubov spectrum of excitation. We obtain it by
seeking the dispersion relation between the frequency $\omega$ and the wave
number $k$ of the linear perturbations around homogenous state of densities 
$\rho_j=|\Psi_j|^2$ respectively.

\begin{equation}
( \omega^2-\frac{k^2}{m_1}(a_1\rho_1+\frac{\hbar^2k^2}{m_1}))
( \omega^2-\frac{k^2}{m_2}(a_2\rho_2+\frac{\hbar^2k^2}{m_2}) ) 
=\frac{g^2 \rho_1 \rho_2 k^4}{m_1 m_2} 
\label{dispersion}
\end{equation}

For uncoupled condensates $g=0$, we retrieve the Bogoliubov spectrum for each 
condensate. The condition for linear stability against demixing ($\omega$ real for
all $k$) leads to the same criterion $a_{1} a_{2} > g^2$.  The coupled equations (\ref{eq1}) are
Galilean invariant and one can thus consider the flow of both condensates at the same constant
velocity through Galilean boosts of the wave-functions. Moreover, notice that for $g=0$, the uncoupled 
eq.  (\ref{eq1}) are Galilean invariant separately so that one can consider a relative constant 
flow between each species.  For weak coupling one can then generalize this  property and thus 
consider the relative flow of one species with respect to the other one. If the condensates are
  homogenous, such flow remains an exact solution of the equations and the model allows an extra 
  "superfluid" property, that is the two species can flow one into each other without dissipation. 
  For inhomogenous condensates like those containing vortices for instance, the interaction between the two species
  generates a friction force and the vortex dynamics is affected by the presence of the two species. 
  The goal of the present paper is precisely to exhibit such motion for simple cases.

From the coupled equations (\ref{eq1}) the vortices bear a double 
integer index, denoting the numbers of phase winding of each 
wavefunction around the core of the vortex. 
The solution of equations (\ref{eq1}) for a vortex $(n_{1}, n_{2})$ 
is of the form $\Psi_{j} = e^{-i E_{j}t} e^{i n_{j}\theta} 
\chi_{j}(r)$, where $(r,\theta)$ are the polar coordinates in the 
plane perpendicular to the vortex axis, and where 
$$\hbar E_{j}= a_{j} \rho_{j} + g \rho_{j+1}.$$
The real functions $\chi_{j}(r)$ are 
solutions of two coupled  ordinary differential equations: 
\begin{equation}
    \hbar E_{j} \chi_{j} = 
    -\frac{\hbar^2}{2m_{j}}\left(\chi_{j}''+ \frac{1}{r}\chi_{j}' + 
    n_{j}^2\,\chi_{j}\right) + a_{j} \chi_{j}^3 + 
    + g  \chi_{(j+1)}^2  \chi_{j}
\label{eq2}
\end{equation}
where the primes stand for the derivative along the radius $r$
($\chi_{j}'=d\chi_{j}/dr$).  The asymptotic condition are that, at $r$
very large, $\chi_{j}$ tends to $\rho_{j}^0$, the uniform density of
the species $j$, although $\chi_{j}$ behaves like $r^{n_{j}}$ at $r$
small.  We restrict later our study on the two dimensional case although 3D dynamics
should reveal interesting behavior (unzipping, Kelvin waves...) and is postponed to
further work.
We assume that the multiply charged vortices (at least one of
the $ n_{j}>1$) are unstable and decompose into separated single
charged vortices as it is the case in general for uncoupled
condensates\cite{aronson,jossBEC}.  Moreover, because of the coupling
via the term proportional to $g$ in the original equations, the
vortices of composite index (both $n_{j}=\pm1$) can be stable, with a joint zero
at the same location.  However this is not at all certain in general 
and such composite vortex can decompose into one single charged vortex 
in each condensate, not located at the same position.
The interaction between vortices belonging to different species, like
vortices $(0,1)$ and $(1,0)$ are short ranged, because it depends on
the density distribution near the vortex core (the interaction between
vortices of the same species are long ranged, because of the velocity
field decaying as $1/r$ at large distances from the core)\cite{pismen}.  A
reasonable guess is to assume that for $g$ negative the vortices
of different species attract each other (whatever their relative sign), although their interaction is
repulsive for $g$ positive.  This is based upon the fact that the
`interaction energy' is, in a first approximation (that is for small $g$), represented by the
integral of $g (\rho_{1}-\rho_1^0)(\rho_{2}- \rho_2^0)$, positive (and
repulsive) for $g$ positive and negative and attracting for $g$
negative.  However, because of the Hamiltonian structure of the dynamics, this instability 
is very slow since it manifests through radiation coming from the vortex acceleration\cite{klyat}.
Our numerics is in complete agreement with this point: we observe that the
two vortices stand at the same position for negative $g$ while they describe 
a slow outward spiraling relative motion for positive $g$.

We introduce here a convenient dimensionless version of the model. 
Rescaling space and time by the factors $(m_1m_2a_1a_2)^{1/4}/\hbar$ 
and $(a_1a_2)^{1/2}/\hbar$ respectively we obtain the following set of
two coupled equations:

\begin{equation}
    i\frac{\partial \Psi_{j}}{\partial t} = 
    -\frac{\alpha_j}{2}\nabla^2 \Psi_{j} + \beta_j |\Psi_{j}|^2  \Psi_{j} 
    + g  |\Psi_{(j+1)}|^2  \Psi_{j}
\label{eqless}
\end{equation}

with $\alpha_1=1/\alpha_2=\sqrt{m_2/m_1}$ and $\beta_1=1/\beta_2=
\sqrt{a_1/a_2}$. For uncoupled condensates ($g=0$), we note that the vortex 
solution $\Psi_j^0({\bf r})$ for each 
condensate is determined by a single function $f$:
\begin{equation}
\Psi_j^0({\bf r})=\sqrt{\rho_j^0}f(\sqrt{\frac{\beta_j\rho_j^0}
{\alpha_j}}r)e^{i \epsilon_j(\theta-\beta_j \rho_j^0 t)} 
\label{solvort}
\end{equation}
where $\epsilon_j=\pm 1$ describes the sign of the circulation of each vortex and 
with $f$ real solution of the equation\cite{ginz}:

$$ -\frac{1}{2} \left( f''(r)+\frac{f'}{r}-\frac{f(r)}{r^2}\right)+(f(r)^2-1)f(r)=0 $$

Consider now the effect of a flow on a composite 
vortex, when $g$ is negative (that is when this composite 
vortex is stable). Because of the Galilean invariance of the coupled equations
we have only to study the relative flow of one species (say $1$) with respect to the other one at
constant speed (say along direction $1$, ${\bf v_1}=v_1{\bf e_1}$).
The numerics shows that
at low speed, the single composite vortex splits first into two 
vortices, a $(0,1)$ vortex and a $(1,0)$ vortex. Then the 
two vortices move together at a speed that is a fraction (explanation below) of the speed of the 
moving species. At higher 
speeds, the vortices split apart, one being carried by the fluid of 
the same species, the other one remaining immobile, unaffected by the 
velocity of the other species. 

To understand this, the simplest thing is to solve the original 
equations (\ref{eq1}) by perturbation, assuming the coupling term to 
be small as well as the uniform velocity of species $1$, species $2$ 
being motionless. The zeroth order solution is a pair of $(1,0)$  
and $(0,1)$ vortices, the first one being located at 
${\bf{r_{1}}}(t)$, the other at ${\bf{r_{2}}}(t)$. Without flow speed 
and without interactions both vortices remain where they are. As speed of species $1$ and the 
interaction is turned 
on, one finds by expansion in powers of a unique small parameter, the 
velocity and the strength of the interaction, the time dependent 
solution of the equations. The equation of motion for  ${\bf{r_{1}}}(t)$ 
and  ${\bf{r_{2}}}(t)$ follows from a solvability condition in this 
expansion. The technical details of this derivation will be explained 
in a coming article, but the principle is the same as the one used in 
similar problems, and that is thoroughly exposed in the book by 
Pismen\cite{pismenbook}. There is a solvability condition in the 
expansion because the ground solution is invariant under translation 
of the position of each vortex. At the first order, one has to solve 
a linear equation for the perturbation to the basic solution, with an 
inhomogeous term coming both from the external speed of species $1$ 
and from the interaction. This linear equation has, in general, no 
solution because the homogeneous piece has a non trivial kernel. 
Imposing that the expansion can be done amounts to impose the 
orthogonality (in the sense of a scalar product in a function space) of the 
inhomogeneous term with the kernel of the 
homogeneous equation. This yields a pair of coupled equations for the 
time derivative of the two vortex positions. They read: 
\begin{equation}
\rho_1^0 \epsilon_1 \frac{d{\bf{r_{1}}}(t)}{dt} = \rho_1^0 \epsilon_1 {\bf{v_{1}}} + \frac{g}{2\pi} {\bf e}_z \times \frac{d V(r_{12})}{d{\bf r_{12}}}
    \label{eq2}
    \end{equation}
    and 
    \begin{equation}
\rho_2^0 \epsilon_2 \frac{d{\bf{r_{2}}}(t)}{dt} = - \frac{g}{2\pi} {\bf e}_z \times \frac{d V( r_{12})}{d{\bf r_{12}}}
	\label{eq3}
	\end{equation}   
where ${\bf r_{12}}={\bf r_2-r_1}$, $\times$ corresponds to the vector product, ${\bf e_z}$ is the unit vector perpendicular to the 2D plane and $g \cdot V(|{\bf r_{12}}|)$
is the potential energy of interaction between the two vortices,
depending on the norm of ${\bf r_{12}}$ only~:

$$ V(|{\bf r_{12}}|)= \int d{\bf r}(|\psi_2^0({\bf r-r_{12}},t)|^2-\rho_2^0)
(|\psi_1^0({\bf r},t)|^2-\rho_1^0) $$

Using the vortex profile (\ref{solvort}), the equation of motion simplifies into the set of equations:

\begin{eqnarray}
\frac{d{\bf r_{12}}}{dt}= -{\bf v_1}&-&\frac{g\epsilon_1\epsilon_2}{2\pi}(\epsilon_1 
\rho_1+\epsilon_2\rho_2) {\bf e}_z \times \frac{d U( r_{12})}{d{\bf r_{12}}} \nonumber \\
\rho_1^0 \epsilon_1 \frac{d{\bf r_{1}}}{dt}+\rho_2^0 \epsilon_2 \frac{d{\bf r_{2}}}{dt}&=&\rho_1^0 
\epsilon_1 {\bf v_1}
\label{eqmotion}
\end{eqnarray}

the first one for the relative motion between the vortices, the second one giving momentum conservation. 
Moreover the rescaled potential $U$ is defined through the function $f$ only:

$$ U(|{\bf r_{12}}|)= \int d{\bf r}(f^2(\sqrt{\frac{\beta_2\rho_2^0}{\alpha_2}}|{\bf r-r_{12}}|)-1)
(f(\sqrt{\frac{\beta_1\rho_1^0}{\alpha_1}}r)-1) $$

The equations of motion  keep the Hamiltonian structure of the coupled G-P equations:
\begin{equation}
\rho_j \epsilon_j \frac{d{\bf{r_j}}(t)}{dt} = -{\bf e_z} \times \frac{\delta {\cal H}}{\bf \delta r_j}
\label{hamil}
\end{equation}

with $ {\cal H}=\rho_1\epsilon_1 {\bf e_z}\cdot ({\bf  v_1 \times r_1 }) + \frac{g\rho_1\rho_2}{2\pi} U(r_{12})$. 

However, wave radiations coming from any accelerated motion of the vortices have to be added to
the dynamics. To account for this dissipative effects (for the vortex motion only, the full set of equations 
being still Hamiltonian), one needs to estimate the radiative terms coming from non uniform
vortex motions. Such complicated calculations have been done for a pair of corotating vortices in
the G-P model and it shows that the dynamics slowly deviates
from the Hamiltonian dynamics, with decreasing value of the energy\cite{klyat,pismenbook}. 
The stability of the composite vortex at zero velocity (stability for negative $g$ only) relies on
 this argument. Moreover, this effect disappears for the non radiating equilibrium states moving at 
 constant speed. They can thus be determined by the Hamiltonian dynamics, their stability being
 determined using this argument on the radiative losses. The trajectories of the Hamiltonian system 
 follow:

\begin{equation}
 \frac{g\epsilon_1\epsilon_2}{2\pi}U(r_{12})-\frac{v_1}{\epsilon_1 \rho_1+\epsilon_2\rho_2 } y_{12}=K
 \label{traj}
 \end{equation}
 
 where $K$ is a motion constant deduced from the Hamiltonian dynamics (\ref{hamil}). 
  
An amazing consequence of the equations of motion 
(\ref{eq2}, \ref{eq3}) is that there is a possible equilibrium solution 
at constant speed of the two vortices, such that the force of 
interaction is balanced by a kind of Kutta-Joukovsky force on each 
vortex. When this is possible, the joint velocity of motion is:
 
$$ \frac{\epsilon_1\rho_1}{\epsilon_1\rho_1+\epsilon_2 \rho_2}{\bf v_1},$$
a simple looking 
result. However this joint drift of the two vortices cannot happen if 
the flow speed is too large. From the equations (\ref{eq2}, 
\ref{eq3}), we obtain the following relation between $ {\bf v_1}$, ${\bf r_{12}}$ and
the potential if the equilibrium solution exists:

\begin{equation}
{\bf v_1}= - \frac{g\epsilon_1\epsilon_2}{2\pi}(\epsilon_1 
\rho_1+\epsilon_2\rho_2) U'(r_{12}) {\bf e}_z \times \frac{{\bf r_{12}}}{r_{12}} .
\label{vitfin}
\end{equation}

\begin{figure}
a)\includegraphics[width=6.cm]{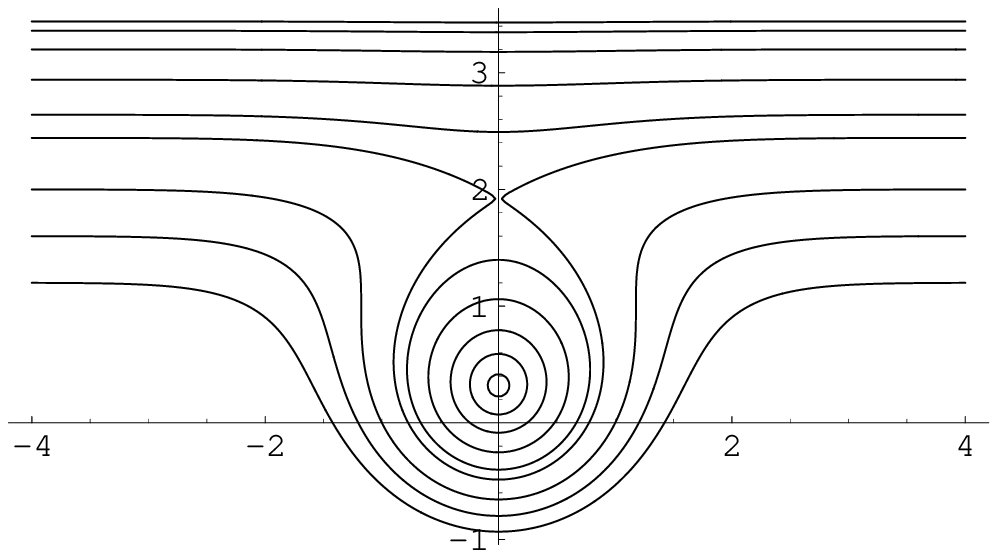}
b)\includegraphics[width=6.cm]{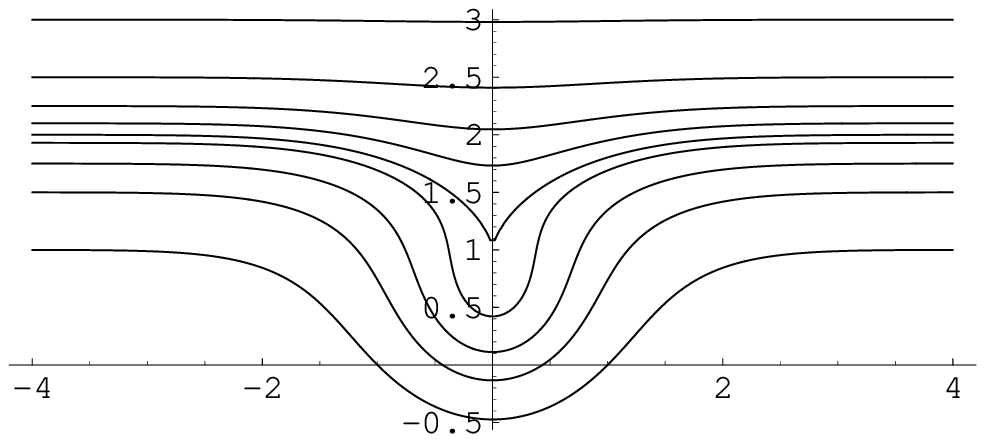} c)\includegraphics[width=6.cm]{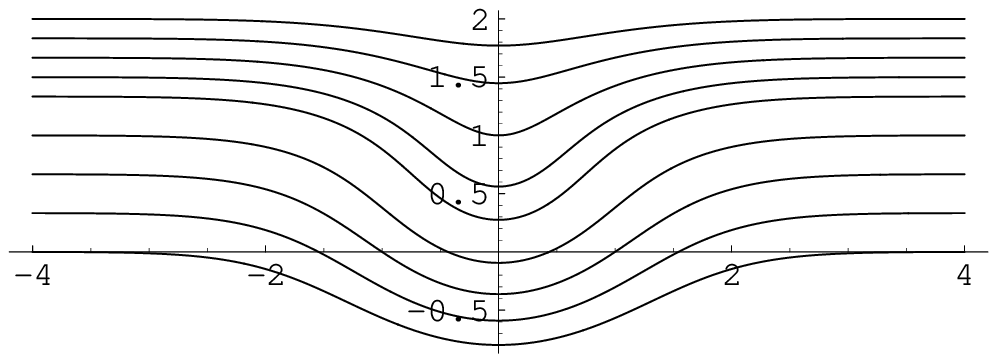} 
\caption{Trajectories of the relative vortex motion for different velocities a) $v_1< v_m$, b) $v_1=v_m$ and c) $v_1>v_m$, neglecting the radiative effects. This trajectories correspond to constant value
of the Hamiltonian.} 
\label{diagramm}
\end{figure}

From this equation, we deduce first that the separation vector  ${\bf r}_{12}$ is orthogonal to the imposed 
velocity $ {\bf v_1}$ and that such a solution can only be found for low enough velocity.
Indeed, the interaction potential has a monotonic behavior with zero derivative both at zero distance 
and at infinity, so that the derivative has a maximum in between which determines as well the 
maximum value $v_m$ of the velocity defined in eq. (\ref{vitfin}). Thus two solutions exist for velocity smaller than this maximum.
The two solutions collapse for the maximum velocity and no more equilibrium positions exist above.
Figure (\ref{diagramm}) shows the Hamiltonian structure of the dynamics for the four cases $v_1=0$,
$v_1<v_m$, $v_1=v_m$ and $v_1>v_m$. The stability of the equilibrium solutions depends then
also on the
sign of $g$. For $g>0$ every solution is linearly unstable and no steady flow can be obtained. For
 $g<0$, we have a
saddle-node bifurcation structure: at low speed when we have two equilibrium positions, one of the
 solution is linearly  stable (the one of smaller $r_{12}$) and the other one unstable. For 
 $v_1=0$, the trajectories are circular and depending on the sign of $g$ the vortex dynamics
 corresponds to a slow converging (diverging) spiraling motion towards (outward) the origin.
Notice also this surprising effect: consider $\epsilon_2=-\epsilon_1$ and $\rho^0_2>\rho^0_1$,
then the constant speed of the two paired vortices is in the opposite direction of the imposed flow velocity!
Such counterflow vortex dynamics has been observed in the numerics as well. Moreover, in the case 
$\epsilon_2=-\epsilon_1$ and $\rho^0_2=\rho^0_1$ the only equilibrium solution is for $ {\bf v_1} =0$.
The experimental consequences of this are simple to state in atomic 
vapors, where one can manipulate the condensate by optical methods 
in particular. It would be interesting also to see if, in superfluid 
mixtures of Helium 4 and 3, there are two kinds of vortices, but it is 
far less easy, but perhaps not impossible, there to move independently 
the two kind of atoms. Another last remark is about the rotating two 
species condensate. With a single species, the equilibrium state is a 
triangular lattice of like sign vortices with a mesh size of order  
$\left(\frac{\hbar}{m\Omega}\right)^{1/2}$, $\Omega$ being the angular frequency. Therefore, if the
two coupled condensate are subject to the same angular speed $\Omega$, each bosonic species 
should in the absence of coupling $g=0$ exhibit a lattice of vortices of mesh 
$\left(\frac{\hbar}{m_j\Omega}\right)^{1/2}$. A non zero coupling induces most probably deformations
of the two lattices. Even in the weak coupling limit, no formal theory seems available for this kind of
situation. Another instance where coupled vortices would be present is the one of nonlinear 
optical fields in the classical approximation\cite{coullet}. There, the role of time in the G-P 
equation is played by the direction parallel to the beam propagation and the Laplacian account
for the 2D variation perpendicular to it. The equations of propagation of two parallel light beams of 
different frequency in the same nonlinear material  are very similar to the coupled G-P equations.
Therefore, we expect in this case the occurence of phenomena roughly similar to the one descibed
here.\\
{\bf Acknowledgements:}\\
It is our pleasure to thank Len Pismen for enlighting discussions.

\ifx\mainismaster\UnDef%
\end{document}
\fi